\PassOptionsToPackage{table,dvipsnames}{xcolor}
\documentclass[conference,10pt]{IEEEtran}
\usepackage[left=0.67in, right=0.68in,top=0.73 in,bottom = 1.05 in]{geometry}
\usepackage{cite}
\usepackage{graphicx,psfrag}
\usepackage{multirow}
\usepackage{array}
\usepackage{setspace}
\usepackage{bbm,mathtools}
\usepackage{amssymb}
\usepackage{color}
\usepackage{enumitem}
\usepackage{stfloats}
\usepackage[flushleft]{threeparttable}
\usepackage{pgfplots}
\usepackage{booktabs}
\usepackage{makecell}
\usepackage[table,dvipsnames]{xcolor}
\pgfplotsset{minor grid style={dotted}}
\pgfplotsset{major grid style={dashed}}
\usepackage{tikz}
\usetikzlibrary{fit,positioning,arrows,shapes,shapes.multipart,calc,arrows.meta,shapes.geometric,shapes.misc}
\usetikzlibrary{decorations.pathreplacing,angles,quotes}
\usepackage[ruled,vlined,linesnumbered]{algorithm2e}
\usepackage[toc,acronym]{glossaries}
\makeindex

\usepackage[labelformat=simple]{subcaption}

\usepackage{grffile}
\pgfplotsset{compat=newest}
\usetikzlibrary{plotmarks}
\usetikzlibrary{arrows.meta}
\usepgfplotslibrary{patchplots}

\usepackage{amsmath,amssymb}
\usepackage{xifthen}
\usepackage{mathtools}
\usepackage{enumerate}
\usepackage{microtype}
\usepackage{xspace}
\usepackage{bm}
\usepackage[T1]{fontenc}
\usepackage{fancyhdr}
\usepackage{lastpage}
\usepackage{bbm}

\newcommand{\mbs}[1]{\pmb{#1}}
\newcommand{\vect}[1]{{\lowercase{\mbs{#1}}}}

\renewcommand{\Re}[1][]{\ifthenelse{\isempty{#1}}{\operatorname{Re}}{\operatorname{Re}\left(#1\right)}}
\renewcommand{\Im}[1][]{\ifthenelse{\isempty{#1}}{\operatorname{Im}}{\operatorname{Im}\left(#1\right)}}

\newcommand{\rv}{\vect{r}}

\newcommand{\phiv}{\vect{\phi}}

\newcommand{\Sc}{{\mathcal S}}

\newcommand{\NN}{\mathbb{N}}

\newcommand{\CN}[1][]{\ifthenelse{\isempty{#1}}{\mathcal{N}_{\mathbb{C}}}{\mathcal{N}_{\mathbb{C}}\left(#1\right)}}

\renewcommand{\P}[1][]{\ifthenelse{\isempty{#1}}{\mathbb{P}}{\mathbb{P}\left[#1\right]}}
\newcommand{\E}[1][]{\ifthenelse{\isempty{#1}}{\mathbb{E}}{\mathbb{E}\left[#1\right]}}
\newcommand{\I}[1][]{\ifthenelse{\isempty{#1}}{\mathbb{I}}{\mathbb{I}\left\{#1\right\}}}
\renewcommand{\det}[1][]{\ifthenelse{\isempty{#1}}{\mathrm{det}}{\mathrm{det}\left(#1\right)}}
\newcommand{\trace}[1][]{\ifthenelse{\isempty{#1}}{{\rm tr}}{\mathrm{tr}\left(#1\right)}}
\newcommand{\rank}[1][]{\ifthenelse{\isempty{#1}}{\mathrm{rank}}{\mathrm{rank}\left(#1\right)}}
\newcommand{\diag}[1][]{\ifthenelse{\isempty{#1}}{\mathrm{diag}}{\mathrm{diag}\left(#1\right)}}
\newcommand{\Cov}[1][]{\ifthenelse{\isempty{#1}}{\mathsf{Cov}}{\mathsf{Cov}\left(#1\right)}}

\newcommand{\defeq}{\triangleq}

\newtheorem{theorem}{Theorem}
\newtheorem{example}{Example}

\newcounter{enumi_saved}
\IfFileExists{MinionPro.sty}{
}{
}

\renewcommand{\rv}[1]{{\mathsf{#1}}}

\newcommand{\of}[1]{^{(#1)}}

\renewcommand{\E}[2][]{\mathbb{E}_{#1}\left[#2\right]}
\renewcommand{\P}[2][]{\mathbb{P}_{#1}\left[#2\right]}

\newcommand{\Bino}{\mathrm{Bino}}

\newcommand{\sub}[1]{_{\mathrm{#1}}}

\renewcommand{\defeq}{=}

\DeclareMathOperator*{\minimize}{minimize}

\newcommand{\hoang}[1]{{#1}} 
\newcommand{\revise}[1]{{#1}} 

\makeatletter
\newcommand{\removelatexerror}{\let\@latex@error\@gobble}
\makeatother

\title{Protocol Design for Irregular Repetition Slotted ALOHA With Energy Harvesting to Maintain Information Freshness \vspace{-.4cm}}

\author{
	\IEEEauthorblockN{Khac-Hoang Ngo\IEEEauthorrefmark{1}, Diep N. Nguyen\IEEEauthorrefmark{2}, Thai-Mai Dinh Thi\IEEEauthorrefmark{3}} 
	\IEEEauthorblockA{\IEEEauthorrefmark{1}Department of Electrical Engineering (ISY), Linköping University, Linköping, SE-58183 Sweden}
    \IEEEauthorblockA{\IEEEauthorrefmark{2}School of Electrical and Data Engineering, University of Technology Sydney, Australia}
    \IEEEauthorblockA{\IEEEauthorrefmark{3}Faculty of Electronics and Telecommunications, VNU University of Engineering and Technology, Hanoi, Vietnam}
	\thanks{The work of Khac-Hoang Ngo was supported in part by the Excellence Center at Linköping – Lund in Information Technology (ELLIIT).}
	\vspace{-.9cm}
}

\newacronym{MAC}{MAC}{multiple access channel}
\newacronym{UMRA}{UMRA}{unsourced massive random access}
\newacronym{SIMO}{SIMO}{single-input multiple-output}
\newacronym{SISO}{SIMO}{single-input single-output}
\newacronym{iid}{i.i.d.}{independent and identically distributed}
\newacronym{ML}{ML}{maximum likelihood}
\newacronym{PEP}{PEP}{pair-wise error probability}
\newacronym{LLR}{LLR}{log-likelihood ratio}
\newacronym{SNR}{SNR}{signal-to-noise ratio}

\newacronym{AoI}{AoI}{age of information}
\newacronym{AVP}{AVP}{age-violation probability}
\newacronym{PMF}{PMF}{probability mass function}
\newacronym{CDF}{CDF}{cumulative distribution function}
\newacronym{SA}{SA}{slotted ALOHA}
\newacronym{IRSA}{IRSA}{irregular repetition slotted ALOHA}
\newacronym{SIC}{SIC}{successive interference cancellation}
\newacronym{PLR}{PLR}{packet loss rate}
\newacronym{DE}{DE}{density evolution}
\newacronym{IoT}{IoT}{Internet of Things}
\newacronym{EH}{EH}{energy harvesting}

\IEEEoverridecommandlockouts
\begin{document}
	
	\maketitle
	\begin{abstract} 
		We investigate an internet-of-things system where energy-harvesting devices send status updates to a common receiver using the irregular repetition slotted ALOHA (IRSA) protocol. Energy shortages in these devices \revise{may lead to} transmission failures that are unknown to the receiver, \revise{disrupting} the decoding process. \revise{To address this issue, we propose} a method for the receiver to perfectly identify \revise{such} failures. Furthermore, we optimize the degree distribution of the protocol to \revise{enhance} the freshness of the status updates. 
        \revise{Our} optimized degree distribution \revise{mitigates the adverse effects} of potential transmission failures. Numerical results \revise{demonstrate} that, despite energy-harvesting \revise{constraints}, IRSA can \revise{achieve} a level of information freshness comparable to systems with unlimited energy.
	\end{abstract}
	
	\section{Introduction} \label{sec:intro}
    To support the sporadic and uncoordinated transmission of a massive number of \gls{IoT} devices, modern random-access protocols have been developed~\cite{Berioli2016NOW} and adopted in commercial applications, such as satellite communication~\cite{ETSI2020DVB}. 
    A prominent example among these protocols is \gls{IRSA}~\cite{Liva2011}, where devices transmit multiple replicas of a packet in randomly chosen slots of a fixed-length frame to create time diversity, and the receiver uses \gls{SIC} to decode. The number of replicas is drawn from a degree distribution.
 
    A common design goal for random-access protocols is to minimize the \gls{PLR} or maximize the throughput
    ~\cite{Liva2011,Ivanov2017,GraelliAmat2018}. However, in emerging time-critical applications, 
    it is also important to ensure the freshness of the receiver's update about the tracked process. Information freshness is captured by the \gls{AoI} metric~\cite{Yates2021AoI}, defined as the time elapsed since the generation of the most recent packet available at the receiver. The \gls{AoI} achieved with random-access protocols has been characterized in various studies, e.g.,~\cite{Yates2017,Yates2020unccordinated,Munari2022_retransmission} for slotted ALOHA and \cite{Munari2020modern,Hoang2021AoI} for \gls{IRSA}.
	
	
	Energy harvesting is a key solution for \gls{IoT} devices with prolonged, low-power operation, \hoang{especially for} remote locations where battery replacement is impractical. It enables the devices to capture and convert ambient energy sources 
    into electrical energy. However, energy harvesting 
    impacts the \hoang{throughput and AoI} performance of random-access protocols. In \gls{IRSA}, packet replicas intended in a frame might not be transmitted due to energy shortages. The effective distribution of the number of transmitted replicas thus becomes different from the designed degree distribution. Furthermore, after decoding a packet, the receiver does not know which of its intended replicas were dropped, and thus cannot proceed with \gls{SIC}. The authors of~\cite{Demirhan2019} propose a method to compute the effective degree distribution, but assume that the receiver knows the position of the dropped replicas. This assumption was also considered in a study of asynchronous \gls{IRSA} in~\cite{Akyildiz2021}. To avoid dropping replicas, the devices can use only the energy available at the beginning of each frame for transmissions during that frame, as considered in~\cite{Haghighat2023}. However, this neglects the potential energy harvested during the frame, and \hoang{thus misses opportunities to deliver status updates, resulting in throughput loss and AoI increase}. 

	
	In this paper, we aim to minimize the average \gls{AoI} for a status-update system where the devices follow the \gls{IRSA} protocol and rely on energy harvesting. 
    We propose a method that allows the receiver to accurately identify the dropped replicas during the \gls{SIC} process. 
    This method relies only on the conventional assumption that the receiver can recognize collision-free slots. \hoang{Our method validates the critical assumption in~\cite{Demirhan2019,Akyildiz2021}, and enables the devices to plan transmissions based on prospective energy harvested in the frame ahead. 
    Using this method and optimized degree distribution, \gls{IRSA} with energy harvesting significantly reduces the average \gls{AoI} compared to slotted ALOHA. It also achieves only a slightly higher average \gls{AoI} than \gls{IRSA} with unlimited energy.} 

	\section{System Model} \label{sec:model}
	We consider a system with $U$ devices delivering timestamped status updates (also called packets) to a receiver through a wireless channel. Time is slotted and each update is transmitted in a slot. We let the slot length be one and \hoang{assume that each packet transmission consumes one unit of energy.} We further assume that time is divided into frames of $M$ slots and devices are frame- and slot-synchronous.	We consider a collision channel model, in which slots containing a single packet (called singleton slots) always lead to successful decoding, whereas slots containing multiple packets lead to decoding failures. As commonly assumed in the literature, the receiver can distinguish between idle slots (containing no packet), singleton slots, and collision slots (containing more than one packet). \hoang{This assumption is typical for the collision channel model and can be achieved based on cyclic redundancy check or hypothesis testing, which are beyond the scope of this paper.}
	
    We assume that each device has a new update in each slot with probability $\alpha$ independently of the other devices. Consequently, each device has a new update in a frame with probability $\sigma = 1 - (1-\alpha)^M$. 
    The average channel load is given by
	$
	G = U \sigma / M 
	$
    devices per slot.
	
	\subsubsection{Energy Harvesting}
	Each device is equipped with a rechargeable battery of capacity $E$ energy units. The devices harvest energy from the environment to recharge their batteries. We follow the energy harvesting model in~\cite{Demirhan2019,Akyildiz2021,Haghighat2023,Ngo24statusupdate}. In each time slot, one energy unit is harvested by each device with probability~$\eta$, independently of the other slots and other devices. 
    If the battery is full, the devices pause harvesting. 
	A packet is transmitted only if there is available energy, i.e., an intended packet is dropped if the battery is depleted. 
	
	\subsubsection{Access Protocol}
	We assume that the system operates according to the \gls{IRSA} protocol. A device may generate more than one update in a frame, but only the latest update is transmitted in the next frame. An active device sends~$\rv{L}$ identical replicas of its latest update in $\rv{L}$ slots chosen uniformly without replacement from the $M$ available slots. Here, $\rv{L}$ is called the degree of the transmitted packet. \hoang{It follows a probability distribution $\{\Lambda_\ell\}_{\ell=\revise{0}}^{\ell_{\max}}$ where $\Lambda_\ell \defeq \P{\rv{L} = \ell}$ and $\ell_{\max}$ is the maximum degree. We write this distribution using a polynomial notation as
	$
		\Lambda(x) = \sum_{\ell=0}^{\ell_{\max}}\Lambda_\ell x^\ell
	$.}
    When $\rv{L} = 0$, the device discards the update, \hoang{resulting} in a packet loss. When $\rv{L} > 0$, some intended replicas, unknown to the receiver, might be dropped due to the lack of energy. 
	
	Upon successfully receiving an update, the receiver can determine the position of its intended replicas. In practice, this can be done by including in the header of the packet a pointer to the position of its replicas, or by letting the devices and receiver share a hash function used to generate the replicas' positions based on the packet payload. 
	In \hoang{the original} \gls{IRSA}, the receiver performs \gls{SIC} decoding \revise{in each frame} as follows. It first seeks a singleton slot, decodes the packet therein, and then attempts to locate and remove its replicas. 
	These steps are repeated until no singleton slot can be found. 
	
	\subsubsection{Performance Metrics} \label{sec:AoI}
 We let $P\sub{e}$ denote the \gls{PLR}, i.e., the probability that a transmitted packet is not successfully decoded after the \gls{SIC} process. The \gls{PLR} is not known in closed form and can be evaluated using simulation and approximated as in~\cite{Ivanov2017,GraelliAmat2018}. The average throughput is given by $G(1-P\sub{e})$ packets/slot.
	We also define the \gls{AoI} for device~$i$ at time $\tau$ as
	$\delta_i(\tau) \defeq \tau - t_i(\tau)$,
	where $t_i(\tau)$ denotes the \revise{generation time} of the last received update from device $i$ as of time $\tau$. Since the \glspl{AoI} of the devices 
    are stochastically equivalent, we focus on a generic device and omit the device index. 
    The \gls{AoI} grows linearly with time and is reset at the end of a frame only when a new update is successfully decoded. 
    The average value of the \gls{AoI} process is given by~\cite[Prop.~2]{Munari2020modern}
    \vspace{-.1cm}
    \begin{equation}
        \bar{\Delta} = \lim_{T \to \infty} \frac{1}{T} \int_{\tau = 0}^T \delta(\tau) {\rm d} {\tau} = \frac{1}{\alpha} + M \Big(\frac{3}{2} + \frac{1}{\xi} - \frac{1}{\sigma}\Big), \label{eq:avgAoI}
        \vspace{-.08cm}
    \end{equation}
    where $\xi = \sigma(1-P\sub{e})$ is the probability that the \gls{AoI} is reset.
    The \gls{AVP}, defined as the probability that the \gls{AoI} value at the end of a generic frame~$j \in \NN_0$ 
    exceeds a certain threshold $\theta$ at steady state, is given by~\cite[Prop.~3]{Munari2020modern} 
	\begin{align} 
		&\zeta(\theta) \defeq \lim_{j\to\infty}\P{\delta(jM) + M > \theta} \notag \\
%
			&=\begin{cases}
				(1-\xi)^{ \lfloor \theta /M \rfloor - 2} \left[1-\frac{1-(1-\alpha)^{1+(\theta \;{\rm mod}\; M)}}{\sigma}\xi \right]\!, &\text{if~} \theta > 2M,\! \\
				1, &\text{otherwise}.
			\end{cases}\label{eq:AVP}
		\end{align}
	
        \vspace{-.1cm}
        We remark that the expressions of the average \gls{AoI}~\eqref{eq:avgAoI} and \gls{AVP}~\eqref{eq:AVP} hold for every frame-slotted protocol where i) a device has a new update in a slot with probability $\alpha$ and transmits the update in the next frame, and ii) the decoding is performed at the end of the next frame and succeeds with probability~$1 - P\sub{e}$.
		\section{\hoang{Mitigating the Impact of Energy Harvesting}}
		\label{sec:realistic_IRSA}
		We refer to the battery level of a device at the beginning of a frame as the {\em initial battery level}. We let the devices adapt the number of intended replicas in a frame to their initial battery level of that frame. For example, with zero initial energy, the device should refrain from intending many replicas.  With initially full battery, the device can transmit at least $E$ replicas. Consider a generic device and let $\rv{B} \in [0:E]$ denote its initial battery level of a generic frame.\footnote{We use $[m:n]$ to denote the set of integers from $m$ to $n$, and $[n] \defeq [1:n]$.} It evolves according to a Markov chain across frames. We denote its steady-state distribution by $\phiv = (\phi_0, \dots, \phi_{E})$ where $\phi_b = \P{\rv{B} = b}$, $b\in [0:E]$. 
        \hoang{To adapt the number of replicas to $\rv{B}$, we vary the conditional probability $\P{\rv{L} = \ell \;\vert\; \rv{B} = b}$ across values of $b$. With a slight abuse of notation, we denote this probability by~$\Lambda_{\ell,b}$.} 
        It follows that the average degree distribution is 
        \begin{equation}
            \Lambda_\ell = \textstyle\sum_{b = 0}^E \phi_b \Lambda_{\ell,b}. \label{eq:avg_degree_dist}
        \end{equation} 
		
		
		\subsection{The Impact of Energy Harvesting}
		When devices rely on harvested energy, they might drop intended replicas due to the lack of energy. In the worst case, all intended replicas are dropped, \hoang{leading} to the following lower bound on the \gls{PLR}.
		\begin{theorem}[\gls{PLR} lower bound]
			\label{th:PLR_lowerbound}
			At steady state, the \gls{PLR} is lower-bounded by
            \vspace{-.15cm}
			\begin{align} \label{eq:PLR_lowerbound}
				&\underline{P\sub{e}} = \notag  \\ 
                &\phi_0 \Bigg( \sum_{y=1}^{M} \eta (1\!-\!\eta)^{y-1}  \sum_{\ell = 0}^{\ell_{\max}} \Lambda_{\ell,0} \frac{(y-1)! (M-\ell)!}{(y-\ell-1)!M!}  + (1\!-\!\eta)^M\Bigg).
			\end{align}
		\end{theorem}
		\begin{proof}See~Appendix~\ref{proof:PLR_lowerbound}.
		\end{proof}
		
		Theorem~\ref{th:PLR_lowerbound} reveals that, with energy harvesting, unless $\eta = 1$, the \gls{PLR} exhibits a positive error floor. 
		With packet drops, the channel becomes similar to a packet erasure channel~\cite{Ivanov2015error}. In our setting, packet erasures are known to the devices but not the receiver. 
        Unknown positions of dropped replicas complicate the \gls{SIC} process, as illustrated in the following example. 
		
		\begin{example}[Impact of unknown packet drops] \label{example}
			 We consider $M=5$ slots and $4$ active devices whose transmission patterns are depicted in Fig.~\ref{fig:IRSA_EH_example}. If no packet is dropped, following~\gls{SIC}, the receiver sequentially decodes devices~$4$, $1$, $3$, and then~$2$. Thus, all packets are decoded.  
			Now assume that some packets are dropped as indicated in Fig.~\ref{fig:IRSA_EH_example}. These packet drops make a supposedly singleton slot (slot~$3$) become idle, and create another singleton slot (slot~$4$). The packet of device~$3$ can thus be decoded in slot~$4$. However, the receiver cannot directly remove its intended replicas since these replicas might not have been transmitted. In fact, the replica in slot~$5$ was dropped. Subtracting it is equivalent to adding a noise that later prevents resolving any other packet in the same slot.  
		\end{example}
		\def\margin{0.05cm}
		\def\Rx{.75cm}
		\def\Ry{.45cm}
		\tikzset{cross/.style={cross out, draw=black, fill=none, minimum height=\Ry-2*\margin, minimum width=\Rx-2*\margin, inner sep=0pt, outer sep=0pt}, cross/.default={2pt}}
		\begin{figure}[t!]
			\centering
			\subcaptionbox{Transmission pattern of active devices under energy harvesting. Crossed cells represent dropped replicas.
            }
			{\scalebox{.86}{\begin{tikzpicture}[font=\footnotesize]
						\draw[black,thin] (0,0) grid  [xstep=\Rx,ystep=\Ry] (5*\Rx,4*\Ry);
						
						\node[align=center] at (-0.8*\Rx,3.5*\Ry) () {device 1};
						\fill[blue!50] (0*\Rx+\margin,3*\Ry+\margin) rectangle (1*\Rx-\margin,4*\Ry-\margin);
						\fill[blue!50] (3*\Rx+\margin,3*\Ry+\margin) rectangle (4*\Rx-\margin,4*\Ry-\margin);
						\node[cross] at (3.5*\Rx,3.5*\Ry) () {};
						
						\node[align=center] at (-.8*\Rx,2.5*\Ry) () {device 2};
						\fill[LimeGreen] (1*\Rx+\margin,2*\Ry+\margin) rectangle (2*\Rx-\margin,3*\Ry-\margin);
						\fill[LimeGreen] (4*\Rx+\margin,2*\Ry+\margin) rectangle (5*\Rx-\margin,3*\Ry-\margin);
						
						\node[align=center] at (-.8*\Rx,1.5*\Ry) () {device 3};
						\fill[orange!90] (1*\Rx+\margin,1*\Ry+\margin) rectangle (2*\Rx-\margin,2*\Ry-\margin);
						\fill[orange!90] (3*\Rx+\margin,1*\Ry+\margin) rectangle (4*\Rx-\margin,2*\Ry-\margin);
						\fill[orange!90] (4*\Rx+\margin,1*\Ry+\margin) rectangle (5*\Rx-\margin,2*\Ry-\margin);
						\node[cross] at (4.5*\Rx,1.5*\Ry) () {};
						
						\node[align=center] at (-.8*\Rx,.5*\Ry) () {device 4};
						\fill[yellow] (0*\Rx+\margin,0*\Ry+\margin) rectangle (1*\Rx-\margin,1*\Ry-\margin);
						\fill[yellow] (2*\Rx+\margin,0*\Ry+\margin) rectangle (3*\Rx-\margin,1*\Ry-\margin);
						\fill[yellow] (4*\Rx+\margin,0*\Ry+\margin) rectangle (5*\Rx-\margin,1*\Ry-\margin);
						\node[cross] at (2.5*\Rx,.5*\Ry) () {};
						
						\node[align=center] at (.5*\Rx,-.4*\Ry) () {slot $\!1$};
                        \node[align=center] at (1.5*\Rx,-.4*\Ry) () {slot $\!2$};
                        \node[align=center] at (2.5*\Rx,-.4*\Ry) () {slot $\!3$};
                        \node[align=center] at (3.5*\Rx,-.4*\Ry) () {slot $\!4$};
                        \node[align=center] at (4.5*\Rx,-.4*\Ry) () {slot $\!5$};
					\end{tikzpicture}
			}}
			\hspace{.05cm}
			\subcaptionbox{Bipartite graph presentation. Solid edges represent transmissions, and dashed edges dropped replicas. 
            }
			{\scalebox{.4}{\begin{tikzpicture}[font=\LARGE]
						\node[circle,fill=black,inner sep=0pt,minimum size=.5cm,label=below:{slot~$1$}] (v1) at (1,0) {};
						\node[circle,fill=black,inner sep=0pt,minimum size=.5cm,label=below:{slot~$2$}] (v2) at (3,0) {};
						\node[circle,fill=black,inner sep=0pt,minimum size=.5cm,label=below:{slot~$3$}] (v3) at (5,0) {};
						\node[circle,fill=black,inner sep=0pt,minimum size=.5cm,label=below:{slot~$4$}] (v4) at (7,0) {};
						\node[circle,fill=black,inner sep=0pt,minimum size=.5cm,label=below:{slot~$5$}] (v5) at (9,0) {};
						
						\node[rectangle,fill=black,inner sep=0pt,minimum size=.5cm,label=above:{device $1$}] (c1) at (1,3.1) {};
						\node[rectangle,fill=black,inner sep=0pt,minimum size=.5cm,label=above:{device $2$}] (c2) at (3.75,3.1) {};
						\node[rectangle,fill=black,inner sep=0pt,minimum size=.5cm,label=above:{device $3$}] (c3) at (6.25,3.1) {};
						\node[rectangle,fill=black,inner sep=0pt,minimum size=.5cm,label=above:{device $4$}] (c4) at (9,3.1) {};
						
						%
						%
						%
						%
						%
						
						\draw[black,thick] (v1) -- (c1);
						\draw[black,dashed] (v1) -- (c4);
						
						\draw[black,thick] (v2) -- (c2);
						\draw[black,thick] (v2) -- (c3);
						
						\draw[black,dashed] (v3) -- (c4);
      
						\draw[black,dashed] (v4) -- (c1);
						\draw[black,thick] (v4) -- (c3);
						
						\draw[black,thick] (v5) -- (c2);
						\draw[black,dashed] (v5) -- (c3);
						\draw[black,thick] (v5) -- (c4);
						
					\end{tikzpicture}
			}}
			\caption{A frame with $M = 5$ slots and $4$ active devices. The first two devices choose degree~$2$ and the last two choose degree~$3$.  
            }
			\label{fig:IRSA_EH_example}
            \vspace{-.55cm}
		\end{figure}
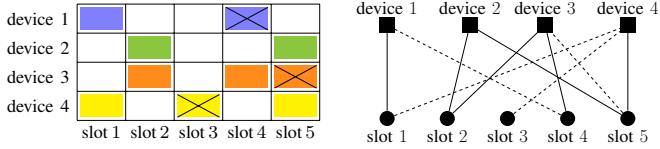
        
        Packet drops alter the transmission pattern of the devices, and thus induce an effective degree distribution of the active replicas. 
        In~\cite{Demirhan2019}, the authors propose a method to derive the effective degree distribution of active replicas. 
        However, they assume that the receiver knows the position of dropped replicas, and do not address how this can be achieved. 
		In the following, we propose methods to circumvent the ambiguity of the dropped replicas' position, and thus validate this critical assumption.
		
		\subsection{Avoid Dropped Replicas}
		The first approach, adopted in~\cite{Haghighat2023}, is to let each device transmit using its energy available at the beginning of the frame only. That is, $\Lambda_{\ell,b} = 0$ for $\ell > b$. In this way, all intended replicas are transmitted, thus the receiver can safely remove all replicas of decoded packets as in conventional \gls{SIC} decoding. Energy harvested in a frame is stored for transmissions in later frames. We refer to this scheme as AVOID.

        We denote by $\Xi_{\ell,b_1}$ the probability that a device spends $\ell$ energy units in a frame given that its initial battery level is~$b_1$. It is computed as follows. If the device is not active or if it is active but choose degree $0$, no energy is spent. Therefore, $\Xi_{0,b_1} = 1 - \sigma + \sigma \Lambda_{0,b_1}$. Otherwise, a degree-$\ell$ active device spends $\ell$ energy units, thus $\Xi_{\ell,b_1} = \sigma \Lambda_{\ell,b_1}$ for $\ell > 0$. 

        We next analyze the evolution of the initial battery level $\rv{B}$. 
		\begin{theorem}[Initial battery level evolution of AVOID] \label{th:schemeA:initial_bat_level}
			The transition probabilities of the initial battery level $\rv{B}$ is given by
			\begin{align} 
				&\!\!\mathbb{P}\big[\rv{B}_{j+1} = b_2 \;|\; \rv{B}_{j} = b_1\big] 
    \notag \\
				&\!\!\!= \begin{cases} \displaystyle
					\sum_{\ell = 0}^{\ell_{\max}} \Xi_{\ell,b_1} \Bino(b_2 \!-\! b_1 \!+\! \ell; M, \eta), \text{~if~} b_1 \!\in\! [0:E\!-\!1],\! \\
					\displaystyle\sum_{\ell = 0}^{\ell_{\max}} \Xi_{\ell,b_1} \displaystyle\Bigg(1 - \sum_{b = 0}^{E - b_1 + \ell - 1}\!\Bino(b - b_1 + \ell; M, \eta)\Bigg), \\
					\hfill \text{if~} b_1 = E.
				\end{cases}\!\!\!\! \label{eq:transition_prob}
			\end{align}
            The stochastic process $\{\rv{B}_{j}\}_j$ is ergodic for every $\ell_{\max} < M$. 
		\end{theorem} 
		\begin{proof}
			See~Appendix~\ref{proof:schemeA:initial_bat_level}.
		\end{proof}
		
        The steady-state distribution $\phiv$ of $\rv{B}$ can be obtained from the transition probabilities~\eqref{eq:transition_prob} by solving the balance equations. Then, we can compute the average degree distribution $\Lambda$ using~\eqref{eq:avg_degree_dist} and analyze the \gls{PLR} using standard analyses of \gls{IRSA} as in, e.g.,~\cite{Liva2011,Ivanov2017,GraelliAmat2018}. This, in turn, allows us to analyze the \gls{AoI} as in~\cite{Munari2020modern,Hoang2021AoI}.

		\subsection{Identify Dropped Replicas}
		\label{sec:scheme_B}
		
		In this approach, we allow the devices to \hoang{plan} more replicas than their initial battery level can support. \hoang{While this may cause dropped replicas, we let} the receiver identify dropped replicas based on its ability to identify singleton slots. Specifically, the receiver forms a list $\Sc_n$ of potentially transmitted packets in each slot $n \in [M]$, called the candidate list. These lists are initially empty. In each \gls{SIC} iteration, the receiver first finds a singleton slot, decodes the packet therein, then adds this packet to the candidate lists in the slots of its intended replicas. Then, for each of these slots, say slot $n$, the receiver tries removing all packets belonging to each possible subset of $\Sc_n$ from the received signal. If after removing any subset, the receiver obtains a singleton slot, all packets in this subset must have been transmitted. They are removed from the received signal in this slot. These steps are repeated until no more singleton slots can be found or a maximum number of iterations has been reached. We illustrate this decoding procedure in Fig.~\ref{fig:schemeB} for the setting in Example~\ref{example}. We refer to this scheme as IDENTIFY. 
  
        \def\Rx{1cm}
            \def\Ry{.6cm}
		\begin{figure}[t!]
			\centering
			\captionsetup[subfigure]{oneside,margin={0.0cm,0cm}}
			\subcaptionbox{\footnotesize Device~$3$'s packet is decoded in slot~$4$ and added to~$\Sc_2$ and~$\Sc_4$.}[.235\textwidth]
			{
                            
				\scalebox{.64}{\begin{tikzpicture}[font=\normalsize]
						\draw[black,thin] (0,0) grid  [xstep=\Rx,ystep=\Ry] (5*\Rx,4*\Ry);
						
						\node[align=center] at (-0.7*\Rx,3.5*\Ry) () {device 1};
						\fill[blue!50] (0*\Rx+\margin,3*\Ry+\margin) rectangle (1*\Rx-\margin,4*\Ry-\margin);
						\fill[blue!50] (3*\Rx+\margin,3*\Ry+\margin) rectangle (4*\Rx-\margin,4*\Ry-\margin);
						\node[cross] at (3.5*\Rx,3.5*\Ry) () {};
						
						\node[align=center] at (-.7*\Rx,2.5*\Ry) () {device 2};
						\fill[LimeGreen] (1*\Rx+\margin,2*\Ry+\margin) rectangle (2*\Rx-\margin,3*\Ry-\margin);
						\fill[LimeGreen] (4*\Rx+\margin,2*\Ry+\margin) rectangle (5*\Rx-\margin,3*\Ry-\margin);
						
						\node[align=center] at (-.7*\Rx,1.5*\Ry) () {device 3};
						\fill[orange!90] (1*\Rx+\margin,1*\Ry+\margin) rectangle (2*\Rx-\margin,2*\Ry-\margin);
						\fill[orange!90] (3*\Rx+\margin,1*\Ry+\margin) rectangle (4*\Rx-\margin,2*\Ry-\margin);
						\fill[orange!90] (4*\Rx+\margin,1*\Ry+\margin) rectangle (5*\Rx-\margin,2*\Ry-\margin);
						\node[cross] at (4.5*\Rx,1.5*\Ry) () {};
						
						\node[align=center] at (-.7*\Rx,.5*\Ry) () {device 4};
						\fill[yellow] (0*\Rx+\margin,0*\Ry+\margin) rectangle (1*\Rx-\margin,1*\Ry-\margin);
						\fill[yellow] (2*\Rx+\margin,0*\Ry+\margin) rectangle (3*\Rx-\margin,1*\Ry-\margin);
						\fill[yellow] (4*\Rx+\margin,0*\Ry+\margin) rectangle (5*\Rx-\margin,1*\Ry-\margin);
						\node[cross] at (2.5*\Rx,.5*\Ry) () {};
						\node at (3.5*\Rx,1.5*\Ry) {\LARGE $\checkmark$};
						
						\node[align=center] at (.5*\Rx,-.4*\Ry) () {slot $\!1$};
                        \node[align=center] at (1.5*\Rx,-.4*\Ry) () {slot $\!2$};
                        \node[align=center] at (2.5*\Rx,-.4*\Ry) () {slot $\!3$};
                        \node[align=center] at (3.5*\Rx,-.4*\Ry) () {slot $\!4$};
                        \node[align=center] at (4.5*\Rx,-.4*\Ry) () {slot $\!5$};
						
						\node[align=center] at (.5*\Rx,4.3*\Ry) () {$\emptyset$};
						\node[align=center] at (1.5*\Rx,4.3*\Ry) () {$\big\{ \colorbox{orange}{\makebox(1.5,3){}} \big\}$};
						\node[align=center] at (2.5*\Rx,4.3*\Ry) () {$\emptyset$};
						\node[align=center] at (3.5*\Rx,4.3*\Ry) () {$\emptyset$};
						\node[align=center] at (4.5*\Rx,4.3*\Ry) () {$\big\{ \colorbox{orange}{\makebox(1.5,3){}} \big\}$};
					\end{tikzpicture}
				}
			}
			\captionsetup[subfigure]{oneside,margin={0.2cm,0cm}}
			\subcaptionbox{\footnotesize The receiver removes device~$3$'s packet from slot~$2$ and obtain a singleton slot. This is not possible in slot~$5$. Device~$2$'s packet is decoded in slot~$2$ and added to~$\Sc_5$.}[.235\textwidth]
			{
				\scalebox{.64}{\begin{tikzpicture}[font=\normalsize]
						\draw[black,thin] (0,0) grid  [xstep=\Rx,ystep=\Ry] (5*\Rx,4*\Ry);
						
						\node[align=center] at (-0.7*\Rx,3.5*\Ry) () {device 1};
						\fill[blue!50] (0*\Rx+\margin,3*\Ry+\margin) rectangle (1*\Rx-\margin,4*\Ry-\margin);
						\fill[blue!50] (3*\Rx+\margin,3*\Ry+\margin) rectangle (4*\Rx-\margin,4*\Ry-\margin);
						\node[cross] at (3.5*\Rx,3.5*\Ry) () {};
						
						\node[align=center] at (-.7*\Rx,2.5*\Ry) () {device 2};
						\fill[LimeGreen] (1*\Rx+\margin,2*\Ry+\margin) rectangle (2*\Rx-\margin,3*\Ry-\margin);
						\fill[LimeGreen] (4*\Rx+\margin,2*\Ry+\margin) rectangle (5*\Rx-\margin,3*\Ry-\margin);
						\node at (1.5*\Rx,2.5*\Ry) {\LARGE $\checkmark$};
						
						\node[align=center] at (-.7*\Rx,1.5*\Ry) () {device 3};
						\fill[orange!90] (4*\Rx+\margin,1*\Ry+\margin) rectangle (5*\Rx-\margin,2*\Ry-\margin);
						\node[cross] at (4.5*\Rx,1.5*\Ry) () {};
						
						\node[align=center] at (-.7*\Rx,.5*\Ry) () {device 4};
						\fill[yellow] (0*\Rx+\margin,0*\Ry+\margin) rectangle (1*\Rx-\margin,1*\Ry-\margin);
						\fill[yellow] (2*\Rx+\margin,0*\Ry+\margin) rectangle (3*\Rx-\margin,1*\Ry-\margin);
						\fill[yellow] (4*\Rx+\margin,0*\Ry+\margin) rectangle (5*\Rx-\margin,1*\Ry-\margin);
						\node[cross] at (2.5*\Rx,.5*\Ry) () {};
						
						\node[align=center] at (.5*\Rx,-.4*\Ry) () {slot $\!1$};
                        \node[align=center] at (1.5*\Rx,-.4*\Ry) () {slot $\!2$};
                        \node[align=center] at (2.5*\Rx,-.4*\Ry) () {slot $\!3$};
                        \node[align=center] at (3.5*\Rx,-.4*\Ry) () {slot $\!4$};
                        \node[align=center] at (4.5*\Rx,-.4*\Ry) () {slot $\!5$};
						
						\node[align=center] at (.5*\Rx,4.3*\Ry) () {$\emptyset$};
						\node[align=center] at (1.5*\Rx,4.3*\Ry) () {$\big\{ \colorbox{orange}{\makebox(1.5,3){}}\big\}$};
						\node[align=center] at (2.5*\Rx,4.3*\Ry) () {$\emptyset$};
						\node[align=center] at (3.5*\Rx,4.3*\Ry) () {$\emptyset$};
						\node[align=center] at (4.5*\Rx,4.3*\Ry) () {$\big\{ \colorbox{orange}{\makebox(1.5,3){}} \ \colorbox{LimeGreen}{\makebox(1.5,3){}}\big\}$};
					\end{tikzpicture}
				}
			}
			\captionsetup[subfigure]{oneside,margin={0.0cm,0cm}}
			\subcaptionbox{\footnotesize The receiver tries removing subsets of $\Sc_5$ from slot~$5$, and succeeds to obtain a singleton slot by removing device~$2$'s packet. Device~$4$'s packet is then decoded in this slot and added to~$\Sc_1$ and~$\Sc_3$.}[.235\textwidth]
			{
				\scalebox{.64}{\begin{tikzpicture}[font=\normalsize]
						\draw[black,thin] (0,0) grid  [xstep=\Rx,ystep=\Ry] (5*\Rx,4*\Ry);
						
						\node[align=center] at (-0.7*\Rx,3.5*\Ry) () {device 1};
						\fill[blue!50] (0*\Rx+\margin,3*\Ry+\margin) rectangle (1*\Rx-\margin,4*\Ry-\margin);
						\fill[blue!50] (3*\Rx+\margin,3*\Ry+\margin) rectangle (4*\Rx-\margin,4*\Ry-\margin);
						\node[cross] at (3.5*\Rx,3.5*\Ry) () {};
						
						\node[align=center] at (-.7*\Rx,2.5*\Ry) () {device 2};
						
						\node[align=center] at (-.7*\Rx,1.5*\Ry) () {device 3};
						\fill[orange!90] (4*\Rx+\margin,1*\Ry+\margin) rectangle (5*\Rx-\margin,2*\Ry-\margin);
						\node[cross] at (4.5*\Rx,1.5*\Ry) () {};
						
						\node[align=center] at (-.7*\Rx,.5*\Ry) () {device 4};
						\fill[yellow] (0*\Rx+\margin,0*\Ry+\margin) rectangle (1*\Rx-\margin,1*\Ry-\margin);
						\fill[yellow] (2*\Rx+\margin,0*\Ry+\margin) rectangle (3*\Rx-\margin,1*\Ry-\margin);
						\fill[yellow] (4*\Rx+\margin,0*\Ry+\margin) rectangle (5*\Rx-\margin,1*\Ry-\margin);
						\node[cross] at (2.5*\Rx,.5*\Ry) () {};
						\node at (4.5*\Rx,.5*\Ry) {\LARGE $\checkmark$};
						
						\node[align=center] at (.5*\Rx,-.4*\Ry) () {slot $\!1$};
                        \node[align=center] at (1.5*\Rx,-.4*\Ry) () {slot $\!2$};
                        \node[align=center] at (2.5*\Rx,-.4*\Ry) () {slot $\!3$};
                        \node[align=center] at (3.5*\Rx,-.4*\Ry) () {slot $\!4$};
                        \node[align=center] at (4.5*\Rx,-.4*\Ry) () {slot $\!5$};
						
						\node[align=center] at (.5*\Rx,4.3*\Ry) () {$\big\{ \colorbox{yellow}{\makebox(1.5,3){}}\big\}$};
						\node[align=center] at (1.5*\Rx,4.3*\Ry) () {$\big\{ \colorbox{orange}{\makebox(1.5,3){}}\big\}$};
						\node[align=center] at (2.5*\Rx,4.3*\Ry) (){$\big\{ \colorbox{yellow}{\makebox(1.5,3){}}\big\}$};
						\node[align=center] at (3.5*\Rx,4.3*\Ry) () {$\emptyset$};
						\node[align=center] at (4.5*\Rx,4.3*\Ry) () {$\big\{ \colorbox{orange}{\makebox(1.5,3){}} \ \colorbox{LimeGreen}{\makebox(1.5,3){}}\big\}$};
					\end{tikzpicture}
			}}
			\captionsetup[subfigure]{oneside,margin={0.2cm,0cm}}
			\subcaptionbox{\footnotesize The receiver can remove device~$4$'s packet from slot~$1$ and obtain a singleton slot. Device~$1$'s packet is then decoded and all devices are resolved.}[.235\textwidth]
			{
				\scalebox{.64}{\begin{tikzpicture}[font=\normalsize]
						\draw[black,thin] (0,0) grid  [xstep=\Rx,ystep=\Ry] (5*\Rx,4*\Ry);
						
						\node[align=center] at (-0.7*\Rx,3.5*\Ry) () {device 1};
						\fill[blue!50] (0*\Rx+\margin,3*\Ry+\margin) rectangle (1*\Rx-\margin,4*\Ry-\margin);
						\fill[blue!50] (3*\Rx+\margin,3*\Ry+\margin) rectangle (4*\Rx-\margin,4*\Ry-\margin);
						\node[cross] at (3.5*\Rx,3.5*\Ry) () {};
						\node at (.5*\Rx,3.5*\Ry) {\LARGE $\checkmark$};
						
						\node[align=center] at (-.7*\Rx,2.5*\Ry) () {device 2};
						
						\node[align=center] at (-.7*\Rx,1.5*\Ry) () {device 3};
						\fill[orange!90] (4*\Rx+\margin,1*\Ry+\margin) rectangle (5*\Rx-\margin,2*\Ry-\margin);
						\node[cross] at (4.5*\Rx,1.5*\Ry) () {};
						
						\node[align=center] at (-.7*\Rx,.5*\Ry) () {device 4};
						\fill[yellow] (2*\Rx+\margin,0*\Ry+\margin) rectangle (3*\Rx-\margin,1*\Ry-\margin);
						\node[cross] at (2.5*\Rx,.5*\Ry) () {};
						
						\node[align=center] at (.5*\Rx,-.4*\Ry) () {slot $\!1$};
                        \node[align=center] at (1.5*\Rx,-.4*\Ry) () {slot $\!2$};
                        \node[align=center] at (2.5*\Rx,-.4*\Ry) () {slot $\!3$};
                        \node[align=center] at (3.5*\Rx,-.4*\Ry) () {slot $\!4$};
                        \node[align=center] at (4.5*\Rx,-.4*\Ry) () {slot $\!5$};
						
						\node[align=center] at (.5*\Rx,4.3*\Ry) () {$\big\{ \colorbox{yellow}{\makebox(1.5,3){}}\big\}$};
						\node[align=center] at (1.5*\Rx,4.3*\Ry) () {$\big\{ \colorbox{orange}{\makebox(1.5,3){}}\big\}$};
						\node[align=center] at (2.5*\Rx,4.3*\Ry) (){$\big\{ \colorbox{yellow}{\makebox(1.5,3){}}\big\}$};
						\node[align=center] at (3.5*\Rx,4.3*\Ry) () {$\emptyset$};
						\node[align=center] at (4.5*\Rx,4.3*\Ry) () {$\big\{ \colorbox{orange}{\makebox(1.5,3){}} \ \colorbox{LimeGreen}{\makebox(1.5,3){}}\big\}$};
					\end{tikzpicture}
			}}
		\caption{Decoding process of IDENTIFY for the \revise{frame} in Fig.~\ref{fig:IRSA_EH_example}. The candidate list of each slot is depicted above the corresponding column. In each step, a packet (with the check mark) is decoded and added to the candidate lists in the slots of its intended replicas. The receiver then tries removing each subset of the candidate lists from the corresponding slots. It succeeds if a singleton slot is obtained.}
		\label{fig:schemeB}
        \vspace{-.4cm}
	\end{figure}
	
	
	With IDENTIFY, the ambiguity of the dropped replicas' positions results in no performance loss, as stated next. 
	\begin{theorem}[Performance guarantee for IDENTIFY] \label{th:schemeB}
		\hoang{Given that the number of \gls{SIC} iterations is unlimited,} following IDENTIFY, the receiver achieves the same \gls{PLR} as if it knows the position of the dropped replicas.
	\end{theorem}
	\begin{proof}
		See Appendix~\ref{proof:schemeB}.
	\end{proof}

    With Theorem~\ref{th:schemeB}, we can proceed to analyze the \gls{PLR} as in the case of known dropped replicas' positions~\cite{Demirhan2019}. 
 
		\hoang{Nevertheless, IDENTIFY has higher complexity and larger decoding delay compared to the case of known dropped replicas' positions. This is due to} the test over all subsets of $\Sc_n$ to identify transmitted packets in slot $n$. The size of $\Sc_n$ is as large as the number of intended replicas in slot~$n$, 
        which can be arbitrarily large when the devices pick slots independently. 
        \hoang{There are some ways to reduce complexity by limiting the size of $\Sc_n$, possibly resulting in a performance loss. First, 
        one can stop decoding in slot $n$ whenever the size of $\Sc_n$ exceeds a certain threshold. Second, to control the maximum size of $\Sc_n$, one can impose a slot-perspective degree distribution with a limited maximum degree, as done in~\cite{Paolini2017GIRSA}. Third, one can include in the header of each packet a pointer to the position of its previously transmitted replicas, so that upon decoding the packet, the receiver can safely remove these replicas.}

\subsection{Numerical Example}
    \begin{figure*} [th!]
		\centering
		\input{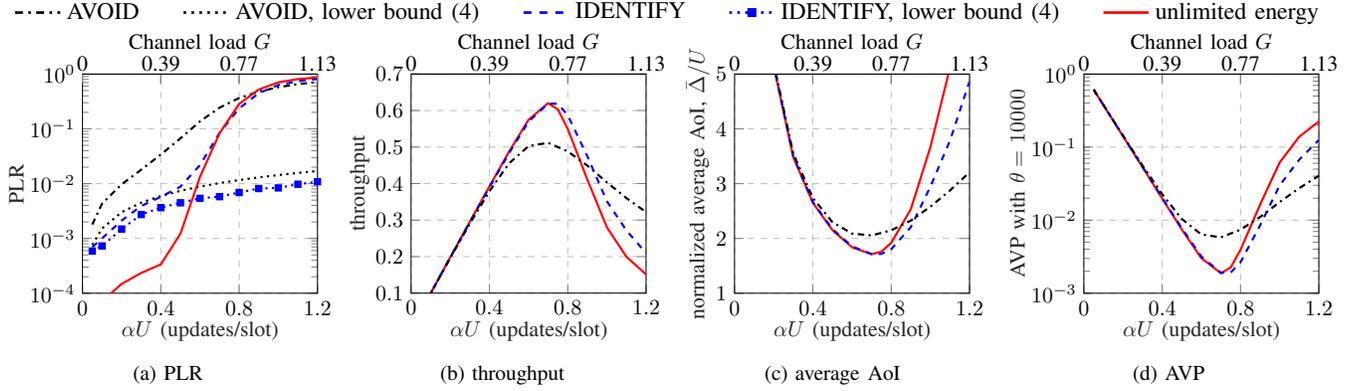}
		\vspace{-.1cm}
        \caption{\gls{PLR}, throughput, average \gls{AoI}, and \gls{AVP} vs. average total number of updates per slot $\alpha U$ for $U = 1000$ devices, $M = 100$ slots, $E = 2$ energy units, and $\eta M = 2$ energy units/frame. For AVOID, we set $\Lambda_b(x) = x^b$, $\forall b \in [0:E]$. For IDENTIFY, we set $\Lambda(x) = x^3$ for all initial battery levels. We also depict the performance achieved with unlimited energy and degree distribution $\Lambda(x) = x^3$.}
		\label{fig:compare_schemes}
        \vspace{-.5cm}
	\end{figure*}

    	We consider a setting with $U = 1000$ devices, frame length $M = 100$ slots, battery capacity $E = 2$ energy units, and average number of harvested energy units per frame $\eta M = 2$. 
	In Fig.~\ref{fig:compare_schemes}, we plot the \gls{PLR}, throughput, average \gls{AoI}, and \gls{AVP} achieved with AVOID with $\Lambda_b(x) = x^b$, $\forall b \in [0:E]$ (i.e., each device uses the entire initial energy to transmit in a frame) and IDENTIFY with degree distribution $\Lambda(x) = x^3$ for all initial battery levels. We also depict the performance achieved with unlimited energy and degree distribution $\Lambda(x) = x^3$. 
    The results are obtained from a Monte-Carlo simulation over $10^5$ frames. 
	For low channel loads, the \glspl{PLR} achieved with AVOID and IDENTIFY are significantly higher than the unlimited energy case due to the potential energy shortage. On the contrary, for high channel loads, both schemes achieve a slightly lower \gls{PLR}, a higher throughput, and lower \gls{AoI} metrics than the unlimited energy case. This is because packet drops reduce collisions. Overall, the throughout and \gls{AoI} metrics achieved with IDENTIFY are similar or even better than the unlimited energy case, suggesting that the devices can operate effectively without a stable power supply. These observations pertain to the considered degree distributions. In the next section, we investigate the optimal degree distributions. 
 
%
 
	\section{Protocol Optimization} \label{sec:optimization}

    We aim to optimize the protocol under the impact of energy harvesting. For concreteness, let us focus on minimizing the average \gls{AoI}; throughput maximization and \gls{AVP} minimization can be done similarly. Specifically, for a fixed $U$, $\alpha$, $E$, $\eta$, and $M$, we consider the optimization
    \vspace{-.1cm}
    \begin{align}
        \minimize_{\Lambda_{\ell, b}, \ell \in [0:\ell_{\max}], b \in [0:E]} &\quad \bar{\Delta} \label{eq:avgAoI_minimization} \\
        {\rm such~that}\quad  \quad &\Lambda_{\ell, b} \in [0,1], \ell \in [0:\ell_{\max}], b \in [0:E], \notag \\
        &\textstyle\sum_{\ell = 0}^{\ell_{\max}} \Lambda_{\ell, b} = 1, \quad b \in [0:E]. \notag
    \end{align}
    For AVOID, we consider an additional constraint $\Lambda_{\ell, b} = 0$ for $\ell > b$. For IDENTIFY, we also consider the case where the degree distribution is not adapted to the initial battery level, i.e., $\Lambda_{\ell, b} = \Lambda_{\ell, b'}$, $\forall \ell, b,b'.$ We evaluate the \gls{PLR} using simulation.  
    We numerically solve~\eqref{eq:avgAoI_minimization} using the Nelder-Mead algorithm~\cite{nelder1965simplex}, a common search method for multidimensional nonlinear optimization. Note that this heuristic method can converge to nonstationary points and is highly sensitive to the initial values. We therefore run the optimization multiple times over $10$ random initializations, and report the best value.  

    In Fig.~\ref{fig:min_avgAoI}, we plot the average \gls{AoI} achieved by the optimized degree distribution for AVOID and IDENTIFY. Here, unless otherwise indicated in each subfigure, we set $U = 1000$ devices, $\alpha U = 1$ update/slot, $M = 100$ slots, $E = 2$ energy units, $\eta = 0.02$ energy units/slot, and $\ell_{\max} = 5$. We also show the performance of i) the slotted ALOHA protocol investigated in~\cite{Ngo24statusupdate}, where the devices immediately transmit their generated update with a probability optimally adapted to their current battery level, and ii) \gls{IRSA} with unlimited energy ($E = \eta = \infty$, $\ell_{\max} = 8$). The following remarks are in order. 
    \begin{itemize}[leftmargin=*]
        
        \item Identifying dropped replicas offers a clear advantage over avoiding them, unless i) $\alpha$ is small (see Fig.~\ref{fig:min_avgAoI}(a)), so that the scarcity of new updates becomes the bottleneck, ii) the battery capacity is high (see Fig.~\ref{fig:min_avgAoI}(b)), so that devices can store sufficient energy to support high degrees, iii) $\eta$ or $M$ is small (see Fig.~\ref{fig:min_avgAoI}(c--d)), so that there is unlikely a new energy harvested during a frame. \revise{For $\alpha U = 1$ and $\eta M = 4$, AVOID results in a $24\%$ higher average \gls{AoI} than IDENTIFY.}

        \item For IDENTIFY, adapting the degree distributions to the initial battery level barely improves over the nonadaptive scheme. Therefore, considering a fixed degree distribution, as done in~\cite{Demirhan2019}, is sufficient in this setting.
        
        \item While the average \gls{AoI} is monotonically reduced as the devices generate new updates more frequently (see Fig.~\ref{fig:min_avgAoI}(a)) or can store/harvest more energy (see Fig.~\ref{fig:min_avgAoI}(b--c)), it is minimized at a certain value of the frame length $M$ (see Fig.~\ref{fig:min_avgAoI}(d)). On the one hand, a small $M$ limits the amount of harvested energy within a frame and thus also limits the feasible number of replicas. On the other hand, a large $M$ leads to a larger initial offset between the generation time and the decoding time of a packet, and a higher number of ignored updates (before the last update generated in a frame). Therefore, $M$ should also be optimized. 

        \item \gls{IRSA} vastly reduces the average \gls{AoI} with respect to slotted ALOHA: the gain is $40.4\%$ for $\alpha U = 1$ and $\eta M = 4$. 

        \item The average \gls{AoI} achieved under energy harvesting is only slightly higher than that with unlimited energy. This confirms that energy harvesting can effectively support \gls{IoT} devices to achieve information freshness.
    \end{itemize}
    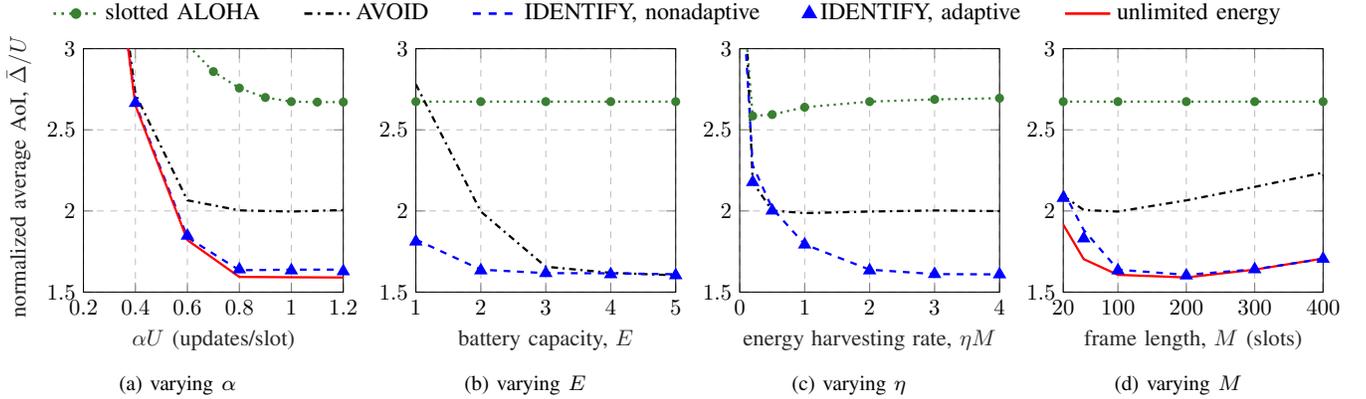
\begin{figure*}
        \centering
                \ref{commonlegend}
        \vskip\baselineskip
        \vspace{-.5cm}
        \subcaptionbox{varying $\alpha$}{
        \begin{tikzpicture}[scale=.85]
		\begin{axis}[%
			width=1.6in,
			height=1.5in,	
			scale only axis,
			unbounded coords=jump,
			xmin=0.2,
			xmax=1.2,
			xlabel style={font=\color{white!15!black}},
			xlabel={$\alpha U$ (updates/slot)},
			ymin=1.5,
			ymax=3,
			yminorticks=true,
			ylabel style={font=\color{white!15!black}},
			ylabel={normalized average \gls{AoI}, $\bar{\Delta}/U$},
			axis background/.style={fill=white},
			title style={font=\bfseries},
            ylabel style={xshift=-.1cm},
			xmajorgrids,
            xminorgrids,
			ymajorgrids,
			yminorgrids,
			legend style={anchor=north, legend columns=-1,legend style={font=\small},/tikz/every even column/.append style={column sep=0.5cm}, draw=none},
            legend to name=commonlegend,
			]
            \addplot [line width = 1,dotted,color=OliveGreen,mark=*,mark options = {solid},mark size = 1.5pt]
			table[row sep=crcr]{%
				1.0e-01 1.1051e+01 \\ 
2.0e-01 6.1058e+00 \\ 
3.0e-01 4.4970e+00 \\ 
4.0e-01 3.7248e+00 \\ 
5.0e-01 3.2895e+00 \\ 
6.0e-01 3.0247e+00 \\ 
7.0e-01 2.8590e+00 \\ 
8.0e-01 2.7571e+00 \\ 
9.0e-01 2.6992e+00 \\ 
1.0e+00 2.6738e+00 \\ 
1.1e+00 2.6709e+00 \\ 
1.2e+00 2.6709e+00 \\ 
			};
            \addlegendentry{slotted ALOHA};
            
            \addplot [line width = 1,dashdotted,color=black]
			table[row sep=crcr]{%
				2.0e-01 5.1834e+0 \\ 
4.0e-01 2.7201e+0 \\ 
6.0e-01 2.0658e+0 \\ 
8.0e-01 2.0037e+0 \\ 
1.0e+00 1.9964e+0 \\ 
1.2e+00 2.0057e+0 \\ 
			};
            \addlegendentry{AVOID};
            
			\addplot [line width = 1,dashed,color=blue]
			table[row sep=crcr]{%
				2.0e-01 5.1567e+0 \\ 
4.0e-01 2.6610e+0 \\ 
6.0e-01 1.8373e+0 \\ 
8.0e-01 1.6348e+0 \\ 
1.0e+00 1.6383e+0 \\ 
1.2e+00 1.6385e+0 \\ 
			}
            ;
            \addlegendentry{IDENTIFY, nonadaptive};

            \addplot [line width = 1,dashed,color=blue,mark=triangle*,mark options={fill=blue,solid},mark size = 2.5pt,only marks]
			table[row sep=crcr]{%
				2.0e-01 5.1615e+0 \\ 
4.0e-01 2.6653e+0 \\ 
6.0e-01 1.8473e+0 \\ 
8.0e-01 1.6398e+0 \\ 
1.0e+00 1.6334e+0 \\ 
1.2e+00 1.6280e+0 \\ 
			};
            \addlegendentry{IDENTIFY, adaptive};

            \addplot [line width = 1,color=red]
			table[row sep=crcr]{%
				2.0e-01 5.1500e+00 \\ 
4.0e-01 2.6500e+00 \\ 
6.0e-01 1.8204e+00 \\ 
8.0e-01 1.5934e+00 \\ 
1.0e+00 1.5915e+00 \\ 
1.2e+00 1.59e+00 \\ 
			};
            \addlegendentry{unlimited energy};
        \end{axis}
        \end{tikzpicture}
        }
        \hspace{-.4cm}
        \subcaptionbox{varying $E$}{
        \begin{tikzpicture}[scale=.85]
		\begin{axis}[%
			width=1.6in,
			height=1.5in,	
			scale only axis,
			unbounded coords=jump,
			xmin=1,
			xmax=5,
			xlabel style={font=\color{white!15!black}},
			xlabel={battery capacity, $E$},
			ymin=1.5,
			ymax=3,
			yminorticks=true,
			axis background/.style={fill=white},
			title style={font=\bfseries},
			xmajorgrids,
            xminorgrids,
			ymajorgrids,
			yminorgrids,
			]
            \addplot [line width = 1,dotted,color=OliveGreen,mark=*,mark options = {solid},mark size = 1.5pt]
			table[row sep=crcr]{%
				1.0e+00 2.6738e+00 \\ 
2.0e+00 2.6738e+00 \\ 
3.0e+00 2.6738e+00 \\ 
4.0e+00 2.6738e+00 \\ 
5.0e+00 2.6738e+00 \\ 
			};
            
            \addplot [line width = 1,dashdotted,color=black]
			table[row sep=crcr]{%
				1.0e+00 2.7806e+00 \\ 
2.0e+00 1.9964e+00 \\ 
3.0e+00 1.6565e+00 \\ 
4.0e+00 1.6187e+00 \\ 
5.0e+00 1.6035e+00 \\ 
			};
            
			\addplot [line width = 1,dashed,color=blue]
			table[row sep=crcr]{%
				1.0e+00 1.8239e+00 \\ 
2.0e+00 1.6383e+00 \\ 
3.0e+00 1.6169e+00 \\ 
4.0e+00 1.6159e+00 \\ 
5.0e+00 1.6100e+00 \\ 
			}
            ;

            \addplot [line width = 1,dashed,color=blue,mark=triangle*,mark options={fill=blue,solid},mark size = 2.5pt,only marks]
			table[row sep=crcr]{%
				1.0e+0 1.8110e+00 \\ 
2.0e+0 1.6334e+00 \\ 
3.0e+0 1.6169e+00 \\ 
4.0e+0 1.6098e+00 \\ 
5.0e+0 1.6035e+00 \\ 
			};
        \end{axis}
        \end{tikzpicture}
        }
        \hspace{-.4cm}
    \subcaptionbox{varying $\eta$}{
        \begin{tikzpicture}[scale=.85]
		\begin{axis}[%
			width=1.6in,
			height=1.5in,	
			scale only axis,
			unbounded coords=jump,
			xmin=0,
			xmax=4,
			xlabel style={font=\color{white!15!black}},
			xlabel={energy harvesting rate, $\eta M$},
			ymin=1.5,
			ymax=3,
			yminorticks=true,
			axis background/.style={fill=white},
			title style={font=\bfseries},
			xmajorgrids,
            xminorgrids,
			ymajorgrids,
			yminorgrids,
			]
            \addplot [line width = 1,dotted,color=OliveGreen,mark=*,mark options = {solid},mark size = 1.5pt]
			table[row sep=crcr]{%
   5.0e-02 3.3243e+00 \\ 
2.0e-01 2.5856e+00 \\ 
5.0e-01 2.5941e+00 \\
				1.0e-0 2.6392e+00 \\ 
2.0e-0 2.6738e+00 \\ 
3.0e-0 2.6875e+00 \\ 
4.0e-0 2.6948e+00 \\ 
5.0e-0 2.6993e+00 \\ 
			};
            
            \addplot [line width = 1,dashdotted,color=black]
			table[row sep=crcr]{%
				5.0e-02 3.6320e+00 \\ 
2.0e-01 2.1777e+00 \\ 
5.0e-01 2.0021e+00 \\ 
1.0e+00 1.9868e+00 \\ 
2.0e+00 1.9964e+00 \\ 
3.0e+00 2.0030e+00 \\ 
4.0e+00 1.9990e+00 \\ 
5.0e+00 1.9960e+00 \\ 
			};
            
			\addplot [line width = 1,dashed,color=blue]
			table[row sep=crcr]{%
				5.0e-02 3.2972e+00 \\ 
2.0e-01 2.2819e+00 \\ 
5.0e-01 2.0167e+00 \\ 
1.0e+00 1.7942e+00 \\ 
2.0e+00 1.6383e+00 \\ 
3.0e+00 1.6109e+00 \\ 
4.0e+00 1.6091e+00 \\ 
5.0e+00 1.6084e+00 \\ 
			}
            ;

            \addplot [line width = 1,dashed,color=blue,mark=triangle*,mark options={fill=blue,solid},mark size = 2.5pt,only marks]
			table[row sep=crcr]{%
				5.0e-02 3.2972e+00 \\ 
2.0e-01 2.1777e+00 \\ 
5.0e-01 2.0021e+00 \\ 
1.0e+00 1.7927e+00 \\ 
2.0e+00 1.6334e+00 \\ 
3.0e+00 1.6109e+00 \\ 
4.0e+00 1.6091e+00 \\ 
5.0e+00 1.6084e+00 \\ 
			};
        \end{axis}
        \end{tikzpicture}
        }
        \hspace{-.4cm}
    \subcaptionbox{varying $M$}{
        \begin{tikzpicture}[scale=.85]
		\begin{axis}[%
			width=1.6in,
			height=1.5in,	
			scale only axis,
			unbounded coords=jump,
			xmin=20,
			xmax=400,
            xtick= {20, 100, 200, 300,400},
			xlabel style={font=\color{white!15!black}},
			xlabel={frame length, $M$ (slots)},
			ymin=1.5,
			ymax=3,
			yminorticks=true,
			axis background/.style={fill=white},
			title style={font=\bfseries},
			xmajorgrids,
            xminorgrids,
			ymajorgrids,
			yminorgrids,
			]

            \addplot [line width = 1,dotted,color=OliveGreen,mark=*,mark options = {solid},mark size = 1.5pt]
			table[row sep=crcr]{%
20 2.6738 \\ 
100 2.6738 \\ 
200 2.6738 \\ 
300 2.6738 \\ 
400 2.6738 \\ 
			};

            \addplot [line width = 1,color=red]
			table[row sep=crcr]{%
				2.0e+01 1.9161e+00 \\ 
5.0e+01 1.7030e+00 \\ 
1.0e+02 1.6063e+00 \\ 
2.0e+02 1.5903e+00 \\ 
3.0e+02 1.6385e+00 \\ 
4.0e+02 1.7067e+00 \\ 
			};
            
            \addplot [line width = 1,dashdotted,color=black]
			table[row sep=crcr]{%
				2.0e+01 2.0944e+00 \\ 
5.0e+01 2.0057e+00 \\ 
1.0e+02 1.9964e+00 \\ 
2.0e+02 2.0653e+00 \\ 
3.0e+02 2.1485e+00 \\ 
4.0e+02 2.2362e+00 \\ 
5.0e+02 0.0000e+00 \\ 
			};
            
			\addplot [line width = 1,dashed,color=blue]
			table[row sep=crcr]{%
				2.0e+01 2.1227e+00 \\ 
5.0e+01 1.8828e+00 \\ 
1.0e+02 1.6383e+00 \\ 
2.0e+02 1.6063e+00 \\ 
3.0e+02 1.6403e+00 \\ 
4.0e+02 1.7052e+00 \\ 
			}
            ;

            \addplot [line width = 1,dashed,color=blue,mark=triangle*,mark options={fill=blue,solid},mark size = 2.5pt,only marks]
			table[row sep=crcr]{%
				2.0e+01 2.0800e+00 \\ 
5.0e+01 1.8304e+00 \\ 
1.0e+02 1.6334e+00 \\ 
2.0e+02 1.6022e+00 \\ 
3.0e+02 1.6403e+00 \\ 
4.0e+02 1.7052e+00 \\ 
			};
        \end{axis}
        \end{tikzpicture}
        } 
        \caption{The minimized average \gls{AoI} vs. $\alpha$, $E$, $\eta$, or $M$. Here, except for the varying parameter, we set $U = 1000$ devices, $\alpha U = 1$ updates/slot, $M = 100$ slots, $E = 2$ energy units, and $\eta = 0.02$ energy units/slot.}
        \label{fig:min_avgAoI}
        \vspace{-.4cm}
    \end{figure*}
	\section{Conclusion} 
    \label{sec:conclusion}
    We studied the performance of \gls{IRSA} when the devices harvest energy to transmit status updates to a receiver. We proposed a method for the receiver to perfectly identify intended replicas that are dropped due to lack of energy.  We also optimized the degree distribution to minimize the average \gls{AoI} metric. 
    We found that the optimal degree distribution should enable the devices to use both energy stored in their battery and potential energy harvested during the upcoming frame. Notably, with energy harvesting, \gls{IRSA} achieves a much lower average \gls{AoI} compared to slotted ALOHA, and a comparable average \gls{AoI} to \gls{IRSA} with unlimited energy. \revise{A future direction is to explore the use of feedback from the receiver.}
	
	\appendix
	
	\subsection{Proof of Theorem~\ref{th:PLR_lowerbound}} \label{proof:PLR_lowerbound}
    \vspace{-.1cm}
	A packet loss is caused by \hoang{either} a packet drop due to the lack of energy or by a decoding failure. By assuming that all transmitted packets are correctly decoded, we lower-bound the \gls{PLR} by the probability that all intended replicas are dropped, denoted by~$\underline{P\sub{e}}$. 
	Consider a degree-$\ell$ packet. Let $\rv{X} \in [M]$ be the index of the slot intended for the last replica of the packet. This random variable has \gls{CDF}
	\begin{equation} \label{eq:CDF_last_replica}
		\P{\rv{X} \le x} = \binom{x}{\ell} / \binom{M}{\ell}  = \frac{x! (M-\ell)!}{(x-\ell)!M!}, \quad x \in [\ell:M],
	\end{equation}
	since there are $\binom{M}{\ell}$ possibilities to choose $\ell$ slots in the frame, and $\binom{x}{\ell}$ possibilities to choose $\ell$ slots among the first $x$ slots. 
	Let $\rv{Y} \in [M]$ denote the index of the slot where the first energy unit is harvested in the frame. The \gls{PMF} of $\rv{Y}$ is given by
 \vspace{-.1cm}
	\begin{align} \label{eq:PMF_first_energy}
		\P{\rv{Y} = y} = \frac{(1-\nu/M)^{y-1} \nu/M}{1-(1-\nu/M)^M}, \quad y \in [M],
	\end{align} 
	where the numerator is the probability that the first energy unit arrives in slot $y$, and the denominator is the probability that at least one energy unit is harvested during the frame. 
	
	All $\ell$ replicas are dropped if the device has no energy in the beginning of the frame,  and either the first harvested energy arrives later than the last intended replica transmission or no energy is harvested during the frame. Therefore, $\underline{P\sub{e}}$ is given~by
	\begin{align}
		&\phi_0 \big(\mathbb{P}\big[\rv{X} < \rv{Y} \vert \rv{B} \!=\! 0\big] + \P{\text{zero harvested energy}}\big) \notag \\
		&= \phi_0 \bigg(\big(1-(1-\eta)^M \big) \sum_{\ell=0}^{\ell_{\max}} \!\Lambda_{\ell,0}\sum_{y=1}^{M} \P{\rv{Y} = y} \P{\rv{X}  < y}  \notag \\ 
        &\qquad \quad + (1-\eta)^M\bigg). \label{eq:tmp428}
	\end{align}
	Finally, by introducing the \gls{CDF} of $\rv{X}$ in~\eqref{eq:CDF_last_replica} and \gls{PMF} of $\rv{Y}$ in~\eqref{eq:PMF_first_energy} into~\eqref{eq:tmp428}, we obtain the expression of $\underline{P\sub{e}}$ in~\eqref{eq:PLR_lowerbound}.
	
	\subsection{Proof of Theorem~\ref{th:schemeA:initial_bat_level}} \label{proof:schemeA:initial_bat_level}
    To show~\eqref{eq:transition_prob}, consider a device that spends $\ell$ energy units in frame~$j$. To jump from battery level $b_1$ to $b_2$ with $b_2 < E$, the device must harvest $b_2 - b_1 + \ell$ energy units.  If at least $E - b_1 + \ell$ units are harvested, the battery is fully recharged, i.e., $b_2 = E$. From this, a direct application of the law of total probability lead to~\eqref{eq:transition_prob}.
    
	We now show that the chain describing $\rv{B}$ is ergodic. Consider an arbitrary state pair $(b_1,b_2)$. To jump from $b_1$ to $b_2$ over $n$ frames, the device could harvest $\rv{H} \le M n$ and spend $\rv{L} \le \ell_{\max} n$ energy units such that $\rv{H} - \rv{L} = b_2 - b_1 \in [-E:E]$. To show that  $E_{j}$ is irreducible, it suffices to show that there exists $n>0$ such that $\P{\rv{H} - \rv{L} = b}$ is strictly positive for every $b \in [-E:E]$. Indeed, if $b \in [0:E]$, this is true because
	\begin{align}
		\P{\rv{H} - \rv{L} = b} &= \P{\rv{H} = \rv{L} + b} \\
		&\ge \P{\rv{H} = \ell_{\max} n + b} \P{
			\rv{L} = \ell_{\max} n} \\
		&> 0, \quad \forall n \ge e/(M - \ell_{\max}),
	\end{align}
	since $\P{\rv{L} = \ell_{\max} n} = \big(\Lambda_{\ell_{\max}}\big)^n > 0$ and $\ell_{\max} n + b$ is in the support $[0:Mn]$ of $\rv{H}$ for every $n \ge b/(M - \ell_{\max})$. For~$b \in [-E:-1]$, it holds in a similar manner that 
	\begin{align}
		\P{\rv{H} - \rv{L} = b} &\ge \P{\rv{H} = \ell_{\max} n + b} \P{\rv{L} = \ell_{\max} n} \\
		&> 0, \quad \forall n \ge -b/\ell_{\max}.
	\end{align}
	Therefore, the Markov chain $\{E_{j}\}_j$ is irreducible. Furthermore, since each state in $[0:E]$ has period $1$, the chain is also aperiodic. Finally, since this irreducible chain is finite, it is also positive  recurrent, and thus ergodic. 
 
	
    \subsection{Proof of Theorem~\ref{th:schemeB}} \label{proof:schemeB}
    It is convenient to consider a graph-based interpretation of the decoding processing of IDENTIFY.
	Consider a bipartite graph where variable nodes~(VNs) represent devices, check nodes~(CNs) represent slots, and VN~$k$ is connected to CN~$s$ by an edge if device~$k$ intends to transmit in slot~$s$. 
    This corresponds to the graph in Fig.~\ref{fig:IRSA_EH_example}(b) with both solid and dashed edges. 
	The receiver initially cannot distinguish between these two edge types. 
    The \gls{SIC} decoding process of IDENTIFY is equivalent to erasure decoding on this graph. Let $\bar{\Sc}_n$ denote the list of revealed edges connected to CN $n$. It contains the edges that connect CN $n$ and the VNs corresponding to the devices in $\Sc_n$. This list is updated after each iteration. The graph-based decoding rules of IDENTIFY are as follows.
    \begin{itemize}[leftmargin=*]
        \item From the VN's perspective, if an edge has been revealed, every other edge connected to the same VN is also revealed and added to the list $\bar{\Sc}_n$ of all corresponding CNs $n$. 
        
        \item From the CN's perspective, an edge connected to CN~$n$ is revealed if $\bar{\Sc}_n$ contains all other edges connected to this CN.  
    \end{itemize}

For a genie-aided decoder that knows the position of the dropped replicas, the \gls{SIC} decoding process is equivalent to erasure decoding on the effective graph that corresponds to the active replicas (i.e., the solid edges) only:
\begin{itemize}[leftmargin=*]
    \item From the VN's perspective, if an edge has been revealed, every other edge connected to the same VN is also revealed.
    
    \item From the CN's perspective, an edge connected to CN~$n$ is revealed if all other edges connected to this CN have been revealed. 
\end{itemize}

We now show that the two graph-based erasure decoding procedures described above are equivalent. 
Specifically, the VN-perspective message update rule is clearly the same in both cases. The CN-perspective message update rule is also equivalent. Indeed, for the genie-aided decoder, an edge connected to a CN~$n$ is revealed if all other active edges connected to the same CN have been revealed. But this is true if and only if these edges have all been included in $\bar{\Sc}_n$. Therefore, after an unlimited number of \gls{SIC} iterations, any active edge that can be revealed in IDENTIFY can also be revealed by the genie-aided decoder, and vice versa. This leads to identical \gls{PLR} of IDENTIFY and the genie-aided scheme.
	
	\bibliographystyle{IEEEtran}
	\bibliography{IEEEabrv,./biblio}

\begin{thebibliography}{10}
\providecommand{\url}[1]{#1}
\csname url@samestyle\endcsname
\providecommand{\newblock}{\relax}
\providecommand{\bibinfo}[2]{#2}
\providecommand{\BIBentrySTDinterwordspacing}{\spaceskip=0pt\relax}
\providecommand{\BIBentryALTinterwordstretchfactor}{4}
\providecommand{\BIBentryALTinterwordspacing}{\spaceskip=\fontdimen2\font plus
\BIBentryALTinterwordstretchfactor\fontdimen3\font minus \fontdimen4\font\relax}
\providecommand{\BIBforeignlanguage}[2]{{%
\expandafter\ifx\csname l@#1\endcsname\relax
\typeout{** WARNING: IEEEtran.bst: No hyphenation pattern has been}%
\typeout{** loaded for the language `#1'. Using the pattern for}%
\typeout{** the default language instead.}%
\else
\language=\csname l@#1\endcsname
\fi
#2}}
\providecommand{\BIBdecl}{\relax}
\BIBdecl

\bibitem{Berioli2016NOW}
M.~Berioli, G.~Cocco, G.~Liva, and A.~Munari, ``Modern random access protocols,'' \emph{Foundations and Trends in Networking}, vol.~10, no.~4, pp. 317--446, Nov. 2016.

\bibitem{ETSI2020DVB}
ETSI, ``{EN 301 545-2 V1.3.1: Digital Video Broadcasting (DVB); Second Generation DVB Interactive Satellite System (DVB-RCS2); Part 2: Lower Layers for Satellite standard},'' Tech. Rep., Jul. 2020.

\bibitem{Liva2011}
G.~Liva, ``Graph-based analysis and optimization of contention resolution diversity slotted {ALOHA},'' \emph{{IEEE} Trans. Commun.}, vol.~59, no.~2, pp. 477--487, Feb. 2011.

\bibitem{Ivanov2017}
M.~Ivanov, F.~Brannstrom, A.~Graell~i Amat, and P.~Popovski, ``Broadcast coded slotted {ALOHA}: A finite frame length analysis,'' \emph{{IEEE} Trans. Commun.}, vol.~65, no.~2, pp. 651--662, Feb. 2017.

\bibitem{GraelliAmat2018}
A.~Graell~i Amat and G.~Liva, ``Finite-length analysis of irregular repetition slotted {ALOHA} in the waterfall region,'' \emph{{IEEE} Commun. Lett.}, vol.~22, no.~5, pp. 886--889, May 2018.

\bibitem{Yates2021AoI}
R.~D. Yates, Y.~Sun, D.~R. Brown, S.~K. Kaul, E.~Modiano, and S.~Ulukus, ``Age of information: An introduction and survey,'' \emph{IEEE Journal on Selected Areas in Communications}, vol.~39, no.~5, pp. 1183--1210, May 2021.

\bibitem{Yates2017}
R.~D. Yates and S.~K. Kaul, ``Status updates over unreliable multiaccess channels,'' in \emph{Proc. {IEEE} Int. Symp. Inf. Theory {(ISIT)}}, Aachen, Germany, Jun. 2017, pp. 331--335.

\bibitem{Yates2020unccordinated}
------, ``Age of information in uncoordinated unslotted updating,'' in \emph{Proc. {IEEE} Int. Symp. Inf. Theory {(ISIT)}}, Los Angeles, CA, USA, 2020, pp. 1759--1764.

\bibitem{Munari2022_retransmission}
A.~Munari, ``On the value of retransmissions for age of information in random access networks without feedback,'' in \emph{Proc. IEEE Glob. Commun. Conf. (GLOBECOM)}, Rio de Janeiro, Brazil, Dec. 2022, pp. 4964--4970.

\bibitem{Munari2020modern}
------, ``Modern random access: {A}n age of information perspective on irregular repetition slotted {ALOHA},'' \emph{{IEEE} Trans. Commun.}, vol.~69, no.~6, pp. 3572--3585, Jun. 2021.

\bibitem{Hoang2021AoI}
K.-H. Ngo, G.~Durisi, and A.~{Graell i Amat}, ``Age of information in prioritized random access,'' in \emph{Proc. Asilomar Conf. Signals, Systems and Computers}, Pacific Grove, CA, USA, Oct. 2021, pp. 1502--1506.

\bibitem{Demirhan2019}
U.~Demirhan and T.~M. Duman, ``Irregular repetition slotted {ALOHA} with energy harvesting nodes,'' \emph{IEEE Trans. Wireless Commun.}, vol.~18, no.~9, pp. 4505--4517, Sep. 2019.

\bibitem{Akyildiz2021}
T.~Akyıldız, U.~Demirhan, and T.~M. Duman, ``Energy harvesting irregular repetition {ALOHA} with replica concatenation,'' \emph{IEEE Trans. Wireless Commun.}, vol.~20, no.~2, pp. 955--968, Feb. 2021.

\bibitem{Haghighat2023}
J.~Haghighat and T.~M. Duman, ``Analysis of coded slotted {ALOHA} with energy harvesting nodes for perfect and imperfect packet recovery scenarios,'' \emph{IEEE Trans. Wireless Commun.}, vol.~22, no.~11, pp. 7424--7437, Nov. 2023.

\bibitem{Ngo24statusupdate}
K.-H. Ngo, G.~Durisi, A.~Munari, F.~L{\'a}zaro, and A.~Graell~i Amat, ``Timely status updates in slotted {ALOHA} networks with energy harvesting,'' \emph{arXiv preprint arXiv:2404.18990}, 2024.

\bibitem{Ivanov2015error}
M.~Ivanov, F.~Br\"{a}nnstr\"{o}m, A.~Graell~i Amat, and P.~Popovski, ``Error floor analysis of coded slotted {ALOHA} over packet erasure channels,'' \emph{IEEE Commun. Lett.}, vol.~19, no.~3, pp. 419--422, Mar. 2015.

\bibitem{Paolini2017GIRSA}
E.~Paolini, G.~Liva, and A.~Graell~i Amat, ``A structured irregular repetition slotted {ALOHA} scheme with low error floors,'' in \emph{Proc. {IEEE} Int. Conf. Communications {(ICC)}}, Paris, France, Jul. 2017, pp. 1--6.

\bibitem{nelder1965simplex}
J.~A. Nelder and R.~Mead, ``A simplex method for function minimization,'' \emph{The computer journal}, vol.~7, no.~4, pp. 308--313, Jan. 1965.

\end{thebibliography}
\end{document}